\documentclass[twocolumn,showpacs,preprintnumbers,amsmath,amssymb]{revtex4}

\usepackage{graphicx}
\usepackage{dcolumn}
\usepackage{bm}
\usepackage{array}
\usepackage{hhline}
\usepackage{subfigure}
\usepackage{amssymb}
\newcommand{\be}{\begin{equation}}
\newcommand{\ee}{\end{equation}}
\newcommand{\bea}{\begin{eqnarray}}
\newcommand{\eea}{\end{eqnarray}}

\begin{document}

\preprint{hep-ph/0506134}

\title{Prediction for the transverse momentum distribution \\ of Drell-Yan dileptons at PANDA \thanks{work
supported by BMBF}}

\author{O.~Linnyk}%
 \email{olena.linnyk@theo.physik.uni-giessen.de}
\author{K.~Gallmeister}
\author{S.~Leupold}
\author{U.~Mosel}
\affiliation{%
Institut f\"ur Theoretische Physik, Universit\"at Giessen, Germany
}%

\date{\today}

\begin{abstract}
We predict the triple differential cross section of the Drell-Yan
process $p\bar p\to l^+l^-X$ in the kinematical regimes relevant
for the upcoming PANDA experiment, using a model that accounts for
quark virtuality as well as primordial transverse momentum. We
find a cross section magnitude of up to 10~nb in the low mass
region. A measurement with 10\% accuracy is desirable in order to
constrain the partonic transverse momentum dispersion and the
spectral function width within $\pm$50~MeV and to study their
evolution with  $M$ and $\sqrt{s}$.
\end{abstract}


\maketitle


\label{model}

\begin{figure}
\begin{center}
\resizebox{0.48\textwidth}{!}{%
  \includegraphics{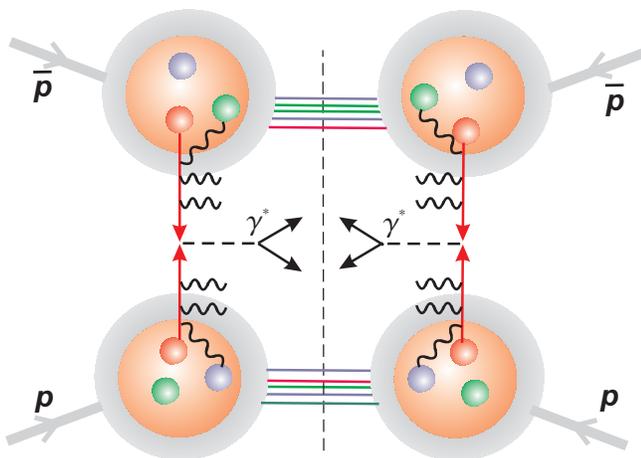}
}
\end{center}
\vspace{-0.5cm} \caption{Relevant initial state interactions that
can build up intrinsic transverse momentum and width of quarks in
the proton in the Drell-Yan process.} \label{diagram}
\end{figure}

Using the model described and tested in an earlier
publication~\cite{paper2}, we calculate the $p_T$-spectrum of the
Drell-Yan pairs in $\bar p p$ collision at the centre of mass
energy $\sqrt{s}=5.5$~GeV, which is an interesting, not yet
experimentally explored kinematical regime. The planned experiment
PANDA~\cite{PANDA} at future GSI facility is to probe the
Drell-Yan process at this energy.

The conventional perturbative QCD (pQCD) approach to the
calculation of hard scattering cross sections assumes that the
partons are collinear and on-shell, {\it i.e.} both the virtuality
and transverse momenta of the quarks and gluons inside hadrons are
neglected and the partonic four-momentum is replaced with it's
light cone projection $(k^+,0,\vec 0)$. This approximation to the
kinematics does not effect the cross section of fully inclusive
processes like DIS, but it leads to unrealistic results for the
cross sections of more exclusive processes. Intrinsic transverse
momentum of quarks has been proven to be important for the
interpretation of various cross sections and
asymmetries~\cite{kT}. The virtuality of quarks and gluons in the
proton has attracted considerable attention recently. As shown
in~\cite{paper2}, by taking into account the virtuality
distribution of partons one can explain the K-factor type
corrections to the transverse momentum distribution of Drell-Yan
pairs. The double differential cross section of the Drell-Yan
process $d \sigma / dM dx_F$ can be well described by pQCD in the
conventional collinear on-shell approach. However, the $p_T$
distribution, {i.e.} the triple differential cross section $d
\sigma / d M dx_F d p_T$ in every order of collinear pQCD is
unphysically sharp. One has either to resum all orders by the
Sudakov method, introducing additional regularization parameters,
or to correct the kinematics by taking into account quark
transverse momentum and virtuality. The latter procedure leads to
a very good description of data already in the leading order of
pQCD for the elementary process~\cite{paper2}. The importance of
accounting for the virtuality of partons in other high energy
processes was stressed, for instance, by O.~Benhar and
V.~R.~Pandharipande~\cite{benhar.pandharipande} and, recently, by
J.~Collins and H.~Jung~\cite{collins.offshell}.

The generalization of the usual parton distribution function,
which depends on the virtuality and $k_T$ of the quark in addition
to it's longitudinal momentum $k^+$, is sometimes called "double
unintegrated parton density" or "parton correlation function". But
the terminology is still not fixed, and one of the reasons is that
the gauge invariant definition of these distributions in terms of
parton field operators is not known yet. It is an open question,
whether factorization still holds in the generalized case, in
which the soft part of the cross section depends on the full quark
momentum $(k^+,k^-,k_T)$. Also, the evolution of these parton
distributions has not been studied so far. Before the
factorization theorem is proven, one can calculate measurable
cross sections using a model that {\em assumes} generalized
factorization. Of course, the predictive power of this model has
to be tested. Comparison of such calculations to the data
in~\cite{paper2} allowed us to fix parameters of a simple
factorized parametrization for the unintegrated quark density in
the proton.  In the present paper, we use the same parameters to
calculate the cross section in the kinematics of the future PANDA
experiment in order to obtain a first estimate for the counting
rates and $p_T$-distributions expected at $\sqrt{s}=5.5$~GeV,
$M=1-4$~GeV, which is on the edge of the non-perturbative regime.
To this aim we also compare with the PYTHIA event
generator~\cite{PYTHIA}.

In~\cite{paper2}, a model to calculate the initial state
interaction (ISI) effects in the Drell-Yan process was proposed.
This model assumes, firstly, the generalized factorization and,
secondly, that the soft part of the cross section can be
approximated by a product of functions of the quark $k^+$
momentum, the transverse momentum, and the virtuality $m^2\equiv
k^+k^- - \vec k \, ^2_T$. We use a Gaussian $k_T$-distribution
\be
\label{D} g(\vec{k}_T) = \frac{1}{4 \pi D^2} \exp \left( -\frac{
\vec{k} \,_{T}^2}{4 D^2} \right) , \ee
so that the mean squared partonic intrinsic transverse momentum
$\langle \vec{k}\,_T  ^2 \rangle = 4 D ^2$.

The authors of~\cite{benhar.pandharipande} pointed out the analogy
between the more conventional collinear approach and the plane
wave impulse approximation (PWIA) of many body theory. Corrections
to the PWIA are due to initial and final state (FSI) interactions.
ISI of active and spectator quarks generate quark virtuality and
intrinsic transverse momentum~(see Fig.~\ref{diagram}).
Calculations in the quark-diquark model show that the partonic ISI
corrections to the cross section of the unpolarized Drell-Yan
process are not small~\cite{brodsky}. However, one has to go
beyond the single gluon exchange approximation of~\cite{brodsky}
in order to study the ISI effects quantitatively. At present, the
quark virtuality distribution cannot be calculated from first
principles QCD. Using analogy to many body theory, we parametrize
the quark virtuality distribution as a Breit-Wigner curve of width
$\Gamma$, which is extracted from data
\be
\label{BW} \mbox{A}(m,\Gamma) =
\frac{1}{\pi}\frac{\Gamma}{m^2+\frac{1}{4}\Gamma^2} . \ee
The exact off-shell kinematics and off-shell sub-process cross
section (at leading order in $\alpha_S$) are used. The
intrinsic-$k_T$ approach is a limiting case of our model at
$\Gamma \to 0$.

The model was tested in \cite{paper2} by comparing the triple
differential cross section $d^3\sigma /d M^2 dp_T^2 d x_F$ of the
processes $p p\to\mu^+ \mu ^- +X$ and $p d\to\mu^+ \mu^- +X$ to
the data of the experiments E866~\cite{E866} and E772~\cite{E772}.
Here, $M$ denotes the invariant mass of the produced lepton pair,
$x_F$ its Feynman variable and $p_T$ its transverse momentum. Both
the shape and magnitude of the observed triple differential cross
sections were well reproduced for reasonable values of $\Gamma
\approx 200$~MeV without a need for a $K$-factor.


\begin{figure}
\begin{center}
\resizebox{0.48\textwidth}{!}{%
  \includegraphics{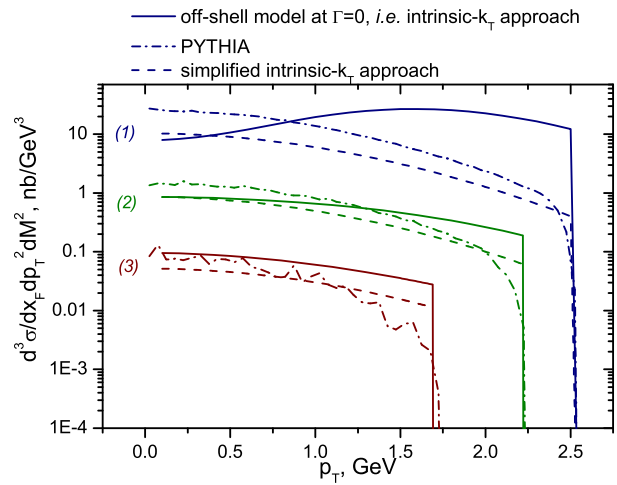}
}
\end{center}
\vspace{-0.5cm} \caption{Comparison of the predictions of
different models: (solid) intrinsic-$k_T$ approach; (dash)
intrinsic-$k_T$ approach with collinear sub-process cross section;
(dash-dot)  PYTHIA generator.
  $x_F=0$, $D=0.5$~GeV, $\sqrt{s}=5.5$~GeV. The lines marked with {\it (1)}
  correspond to  $M=1$~GeV; {\it (2)},~$M=2$~GeV; {\it (3)},~$M=3$~GeV. }
\label{PYTHIA}
\end{figure}

\label{results}

We apply the formalism derived in~\cite{paper2} to predict the
cross section
\be
 \frac{d\sigma}{d\vec p}
\equiv \frac{2}{\pi \sqrt{s}}\frac{d\sigma}{d x_F d p_T ^2}
=\frac{2}{\pi \sqrt{s}} \int \limits _{\mbox{\small bin}} \!
                                \frac{d^3\sigma}{d x_{F} d p_T ^2 dM^2} \, dM^2
\label{triple} \ee
of the process $\bar p p\to l^+l^-X$ in the kinematical regime
planned to be probed by PANDA. We use
\be \nonumber %
\frac{d\sigma}{d x_F d p_T ^2 dM^2} \! = \! \! \! \!
\sum _{\mbox{\small flavors}} \! \! \! (f _{p} \bar f_{\bar p } +
f _{\bar p} \bar f_{p}) \otimes \frac{d \hat \sigma}{d x_F d p_T
^2 dM^2} \otimes \mbox{A}
\bar{\mbox{A}}
, \ee
where $\sqrt{s}$ is the centre of mass energy of the $\bar p p$,
the parton level cross section $d \hat \sigma$ is taken from
\cite{paper2}, the spectral function $\mbox{A}$ is given in
(\ref{BW}), the unintegrated parton distribution $f$ is a product
of (\ref{D}) and the parton distribution functions~\cite{grv}, the
quantities referring to anti-quarks are denoted with bars. We
assume that $\bar \Gamma \! =\! \Gamma$ and $\bar D\! =\! D$.

The two parameters of the model ($D$ and $\Gamma$) should be
extrapolated to the values of $M\! = \! (1-4)$~GeV and $\sqrt{s}\!
=\! 5.5$~GeV relevant for PANDA. In~\cite{paper2}, we have
performed  fits to the Fermilab experiments E866~\cite{E866} and
E772~\cite{E772} in different bins of $M$. From that, we estimate
the parameter $D\approx 0.6\pm 0.18$~GeV. The large uncertainty of
the extrapolation is reflected in our estimate of the parameter
error. The model parameter $D$ should be understood as
representing the summed effect of the transverse motion of partons
inside the nucleon and of the perturbative
corrections~\cite{Alesio.Murgia}: $D^2=D_{\mbox{\small intr}}
^2+D_{\mbox{\small pert}} ^2$. The transverse momentum coming from
higher orders of collinear perturbation theory shows at constant
$\tau\equiv M^2/s$ a linear dependence on $s$ in addition to the
logarithmic dependence on $M^2$~\cite{Altarelli}: $ \mbox{$
\langle p_T^2 \rangle $} _{\mbox{\small pert}} =s \ \alpha_S
\left( M^2 \right) f \left( \tau \right)$.
On the other hand, the non-perturbative $D_{\mbox{\small intr}}$
does not show a strong dependence on $s$. An analysis using the
event generator PYTHIA~\cite{PYTHIA} has shown that the
perturbative corrections constitute only 0.1\%  of the
$p_T$-spectrum of Drell-Yan pairs with $p_T\ne 0$ at PANDA
kinematics. Thus, we set $D \approx D_{\mbox{\small intr}}$.

\begin{figure*}
\begin{center}
\subfigure[$M = 1$~GeV] 
{
         \resizebox{0.48\textwidth}{!}{%
         \includegraphics{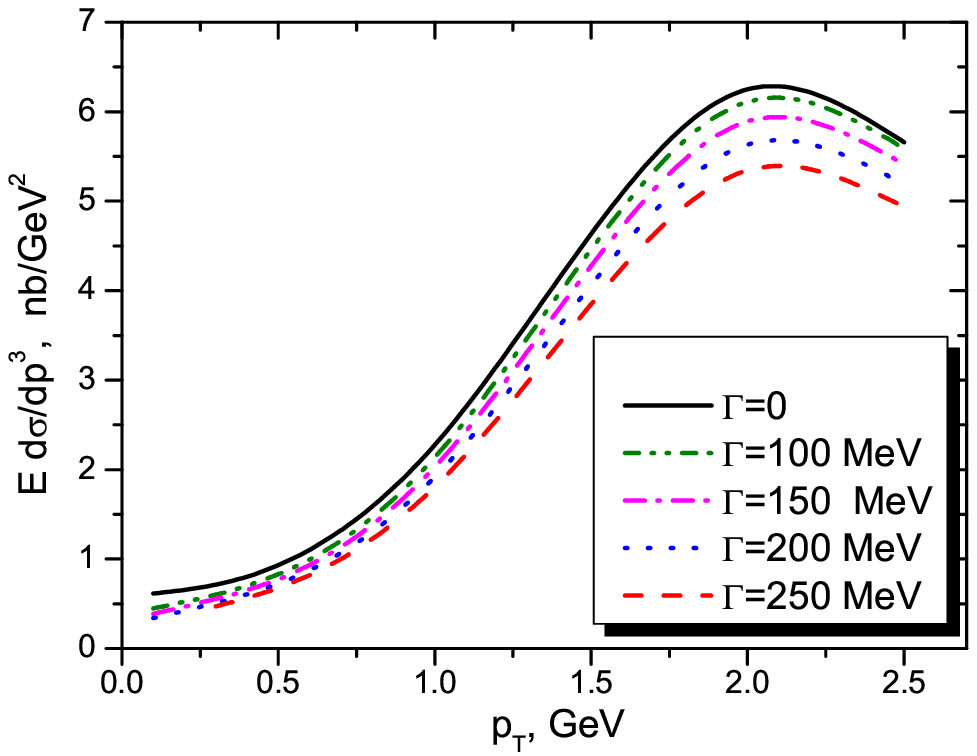}
         }
        \label{Mass1GeV}
} \hspace{-0.5cm}
\subfigure[$M = 2$~GeV] 
{
        \label{Mass2GeV}
         \resizebox{0.48\textwidth}{!}{%
         \includegraphics{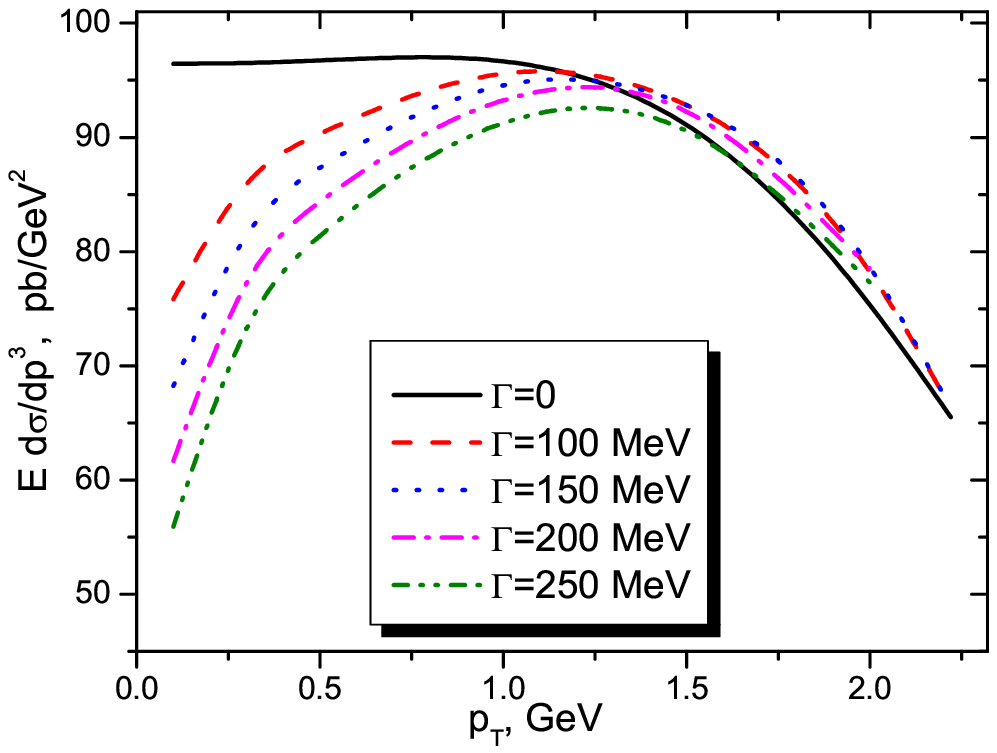}
         }
}
\end{center}
\vspace{-0.5 cm} \caption{ Prediction for the $p_T$-distribution
of Drell-Yan lepton pairs at PANDA. Intrinsic transverse momentum
dispersion $D\!=\! 0.6$~GeV. The solid line is the result of
calculations in the intrinsic-$k_T$ approach (width $\Gamma \!=\!
0$). The other curves are generated with $\Gamma$ in the range
that we have determined by fitting existing Drell-Yan data. Note
that the scale changes from nb in the left figure to pb in the
right one. $\sqrt{s}\! =\! 5.5$~GeV, $x_F\! =\! 0.1$ in all plots.
} \label{prediction1}
\end{figure*}

\begin{figure*}
\begin{center}
\subfigure[$M = 3$~GeV] 
{
         \resizebox{0.48\textwidth}{!}{%
         \includegraphics{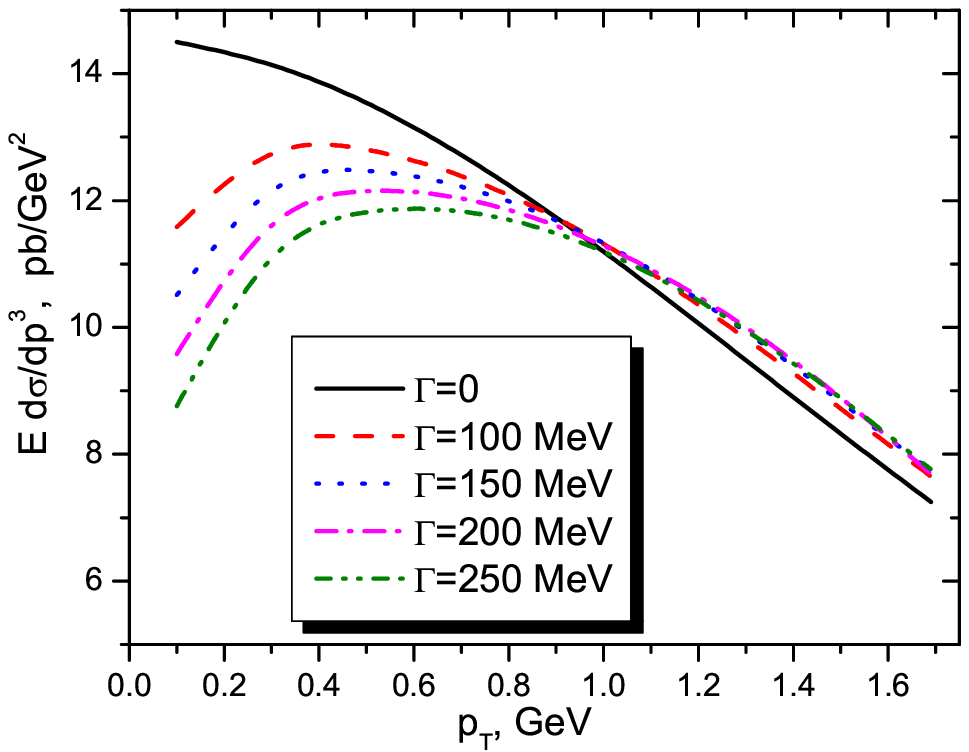}
         }
        \label{Mass3GeV}
} \hspace{-0.5cm}
\subfigure[$M = 4$~GeV] 
{
        \label{Mass4GeV}
         \resizebox{0.48\textwidth}{!}{%
         \includegraphics{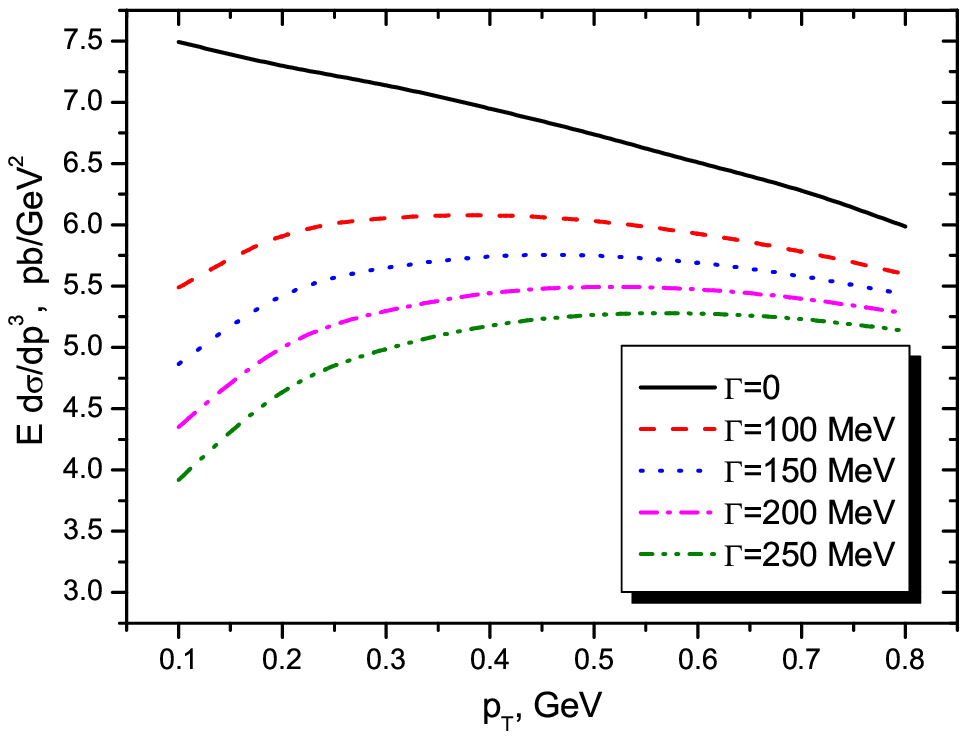}
         }
}
\end{center}
\vspace{-0.5 cm} \caption{Same as Fig.~\ref{prediction1}, but for
higher masses of Drell-Yan
  pairs.}
\label{prediction2}
\end{figure*}

Fig.~\ref{PYTHIA} presents a comparison of the predictions
obtained with three different methods:
1) our model at $\Gamma \! =\! 0$, $D\! =\! 0.5$~GeV, which is
equivalent to the intrinsic-$k_T$ approach (solid lines);
2) the widely used simplified version of the intrinsic-$k_T$
approach, in which the dependence of $\hat \sigma$ on $k_T$ is
neglected (dash line);
3) PYTHIA, taking into account QED and QCD initial state radiation
and intrinsic-$k_T$ (dash-dotted line).
%
The models agree in the overall strength of the cross section
within a factor of 3.
The demand of positive mass of the remnant $M_X$ determines the
maximum $p_T$. In PYTHIA, however, a stronger constraint is
implemented: $M_X\ge2 m_{qq}\approx 1.6$~GeV, where $m_{qq}$ is a
diquark mass. We also used 1.6~GeV as the lower bound for the
remnant mass in the other models for consistency. The dilepton
mass bin width in the simulation was set to 100~MeV. From
Fig.~\ref{PYTHIA}, one observes that the calculations within the
simplified intrinsic approach (dashed line) and within PYTHIA
(dash-dotted line) give similar results. Indeed, the
intrinsic-$k_T$ approach with an approximate collinear sub-process
cross section is implemented in PYTHIA. The difference between the
two curves (dash and dash-dot) at high $p_T$ values are due to a
more detailed modelling of the remnant in the event generator.
Note the qualitative difference of the cross section at the
Drell-Yan pair mass $M\! =1\! $~GeV in the intrinsic-$k_T$
approach (solid lines). In contrast to the higher mass bins, the
peak of the $p_T$-distribution for $M\! =\! 1$~GeV is not at zero
in our model (solid line in Fig.~\ref{PYTHIA}). This behavior
appears at $M \lesssim 2D$, {\it i.e. }in the distribution of the
low virtuality photons, which can be produced by the partonic
transverse motion alone. It is worthwhile to stress that this
drastic change in the $p_T$-dependence of the cross section takes
place for all values of $\Gamma$. An experimental verification of
this effect would constitute a direct test for the transverse
momentum distribution of partons. Note, however, that this effect
appears only if the $k_T$-dependence of the partonic cross section
is not neglected (this is the difference between the solid and
dashed lines in Fig.\ref{PYTHIA}).

\begin{figure}
\begin{center}
\resizebox{0.48\textwidth}{!}{%
  \includegraphics{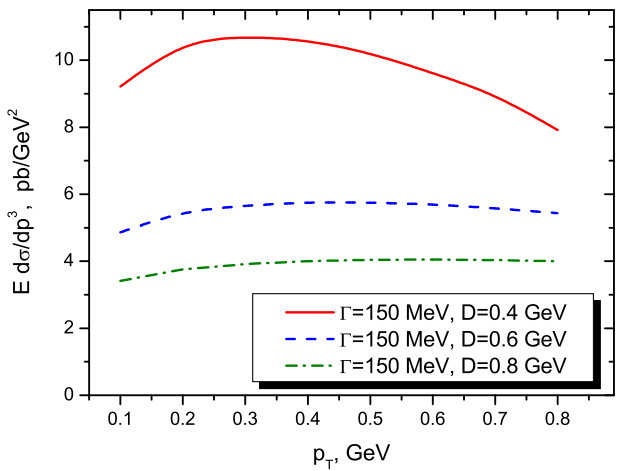}
}
\end{center}
\vspace{-0.5cm} \caption{Variation of the cross section with $D$
at fixed $\Gamma$. $s=32$~GeV$^2$, $x_F=0.1$, $M=4$~GeV.}
\label{Dvary}
\end{figure}

Fig.~\ref{prediction1} and ~\ref{prediction2} show several
theoretical curves for the cross section (\ref{triple})  with
$\Gamma \! = \! (100-250)$~MeV, which is the range determined from
fitting E866 data in~\cite{paper2}. The results of our
calculations in the intrinsic-$k_T$ approach ($\Gamma \! = \! 0$)
are plotted for comparison (solid lines). The evolution of the
spectral function with the hard scale is unknown. It cannot be
directly related to the evolution of the $k_T$-distribution,
because the quark off-shellness depends also on $k^-$. One can see
that the effect of the $\Gamma$ variation on the cross section is
of the order of 10\%.
On the other hand, the variation of the parameter $D$ within
theoretical uncertainty at fixed $\Gamma$ (as presented in
Fig.~\ref{Dvary}) also leads to considerable changes of the cross
section. As was shown in~\cite{paper2}, though, one can do a
double fit and extract both parameters from the same data set.

\label{discussion}

The measurement of this cross section at $\sqrt{s}$ as low as
5.5~GeV will provide essential information on QCD in the regime
where effects beyond leading order and leading twist are expected
to be large~\cite{alphaS2}. The high amplitude of the predicted
cross sections (up to 10 nb) indicates that PANDA at the design
luminosity has a potential to measure the triple differential
unpolarized cross section of $\bar p p\to l^+l^- +X$ with high
statistics and an unprecedented accuracy.

The transverse momentum spectrum of Drell-Yan pairs at low
$\sqrt{s}$ is generated predominantly by the non-perturbative
primordial intrinsic transverse momentum of the partons. On the
other hand, the distribution of the intrinsic $k_T$ in this region
is poorly known. PANDA data will be a valuable input that should
allow one to pin down the quark transverse momentum distribution
in the proton. Again, the $M=1$~GeV mass bin shows the mentioned
shift of the peak towards higher $p_T$.

The presented results also suggest that one can use the future
PANDA data to gain information on the spectral function of partons
bound in the nucleon. Indeed, as we see from
Fig.~\ref{prediction1} and \ref{prediction2}, an experimental
accuracy of 20-30\% would be enough to answer the question of
whether the cross section can be described by a model with
on-shell quarks. The results of the earlier calculations at higher
$\sqrt{s}$ together with the PANDA data can be used also to
extract the dependence of the quark spectral function in the
proton on the hard scales $M$ and $\sqrt{s}$. For this purpose, an
experimental accuracy of at least 10\%  is needed, so that one can
reliably extract the parameter $\Gamma$ in  different mass bins.
Should the accuracy be even better, the data can be used to
investigate the details of the quark spectral function shape.

Work supported by BMBF.



\end{document}